\documentclass[aps,pra,twocolumn,superscriptaddress,epsfig,amsmath,floatfix,showpacs]{revtex4}
\usepackage{epsfig,amsmath}
\usepackage{graphicx}
\usepackage{dcolumn}
\usepackage{bm}

\begin{document}

\title{Nonsequential double ionization: a comparison of classical simulations
and $S$-matrix results}

\author{Phay J. Ho}
\affiliation{Department of Physics and Astronomy, University of Rochester, 600
Wilson Blvd., Rochester, NY 14627, USA}
\affiliation{Max Born Institute for Nonlinear Optics and Short-Pulse Spectroscopy,
Max-Born-Strasse 2A, Berlin 12489, Germany}
\author{X. Liu}
\affiliation{State Key Laboratory of Magnetic Resonance and Atomic and Molecular Physics, Wuhan Institute of Physics and Mathematics,
Chinese Academy of Sciences,
Wuhan 430071,  P.R.China}
\author{W. Becker}
\affiliation{Max Born Institute for Nonlinear Optics and Short-Pulse Spectroscopy, Max-Born-Strasse 2A, Berlin 12489, Germany}

\begin{abstract}%

In a fully classical simulation, we investigate the recollision mechanism of non-sequential double ionization (NSDI) and its manifestation in the end-of-pulse electron momentum distributions. We compare two different electron-electron potentials:  a soft-core Coulombic potential and a Yukawa potential.  We also implement the strong-field approximation (SFA), which is commonly made in quantum-mechanical $S$-matrix calculations, in our classical simulations and study its consequences.   We find that, regardless of the form of the e-e potential, the SFA modifies the momentum distributions and recollision dynamics significantly, but more so for the long-range than for the short-range e-e interaction.  Surprisingly, our classical results, especially those obtained with the Yukawa potential under the SFA, agree well with the results from the $S$-matrix calculations with a contact e-e interaction.  This implies that the recollision dynamics initiated by a quantum tunneling process and a purely classical process do not deviate much from each other. Furthermore, our classical predictions of travel times are consistent with the results from the simple-man model, and the most probable thermalization time is found to be within 0.15 and 0.25 laser periods.

\end{abstract}

\date{\today}

\maketitle

\section{Introduction}

Correlated electron dynamics of atoms and molecules in intense radiation fields remains a challenging and fascinating research topic. One interesting e-e correlated phenomenon is the so-called non-sequential double ionization (NSDI), which has been observed in all inert gas atoms \cite{NSDI-atoms}.  Many efforts have been devoted to understand the e-e correlation underlying this process.  For instance, $S$-matrix calculations \cite{s-matrix-bf, s-matrix-mbi, s-matrix-gp, s-matrix-all} have been used to examine the role of a three-step recollision scenario in NSDI: one electron first tunnels out of the nucleus, collects energy from the laser field  and carries it back to the ion to accomplish the second ionization \cite{corkum}. In addition to relying on this recollision scenario, these calculations make use of the so-called strong-field approximation (SFA) \cite{KFR}, in which the nuclear interaction of the recolliding electron and the outgoing electrons as well as  the interaction of the bound electrons with the laser field are suppressed.

\begin{figure}
\centerline{\includegraphics[width=3.4in]{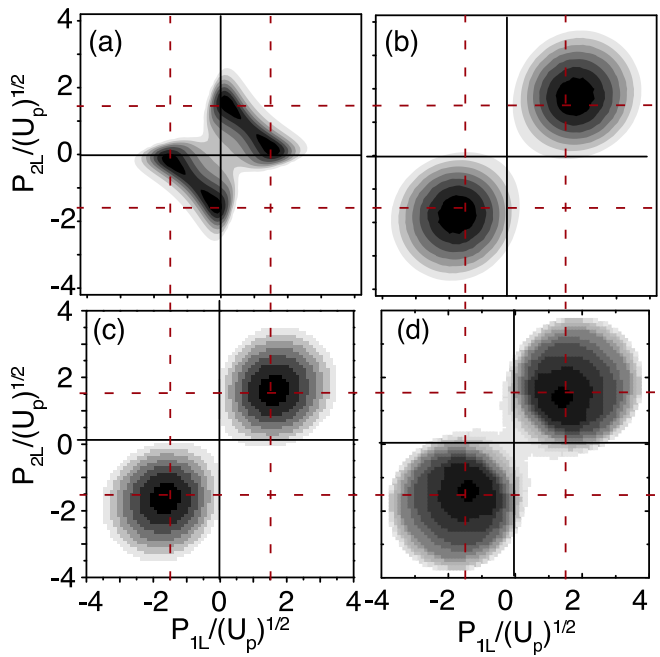}}
\caption{Two-electron longitudinal momentum distributions calculated from the $S$-matrix element that incorporates the tunneling-recollision scenario in the SFA.  The two panels in the top row are obtained from a 3-d calculation with (a) the Coulomb repulsion potential and (b) a contact repulsion potential.  The two panels in the bottom row are obtained from (c) a 3-d and (d) a 2-d statistical quasi-static-tunneling model that employs a contact repulsion potential.  All panels are calculated for a Ti:Sapphire laser with a peak intensity of $I=0.55$ PW/cm$^2$. The vertical and horizontal red dashed lines mark P$_{1L}=\pm 1.5 \sqrt{U_p}$ and P$_{2L}=\pm 1.5\sqrt{U_p}$.  Panels (a), (b) and (c) are taken from \cite{classicalSmatrix2}. }
\label{fig.tunneling}
\end{figure}

A crucial ingredient of the recollision scenario is the form of the electron-electron interaction. In the $S$-matrix calculation with the SFA, this potential, by which the returning electron kicks out the second electron, enters only in first-order Born approximation. Essentially two choices have been examined: the Coulomb repulsion potential \cite{s-matrix-bf,s-matrix-gp,s-matrix-all} and a three-dimensional (3-d) contact potential \cite{s-matrix-mbi,s-matrix-all}.  Figures~\ref{fig.tunneling} (a) and (b) show that the end-of-pulse momentum distributions along the laser polarization axis obtained from these two potentials have few or no NSDI electron pairs with small or opposite momenta.  Apart from this, these two distributions are rather different.  The Coulomb potential predicts that the two NSDI electrons are most likely to have different longitudinal momenta, loosely speaking one fast and one slow electron, whereas the contact potential predicts that both NSDI electrons are mostly like to have about the same longitudinal momentum \cite{s-matrix-all}.  Recently, these features of the contact and the Coulombic potentials have been reproduced remarkably well from 3-d semi-classical calculations based on a statistical model that uses the same rescattering scenario employed in the $S$-matrix calculations \cite{classicalSmatrix1, classicalSmatrix2, classicalSmatrix3}, and a quasistatic tunneling rate \cite{tunneling}. Semiclassical trajectory calculations that start from quantum-mechanical tunneling but thereafter track two-electron classical trajectories have been presented in a series of papers \cite{chenetal}. Even though in these simulations all interactions are Coulomb interactions, the results are rather similar to the $S$-matrix results for contact interactions \cite{s-matrix-mbi,s-matrix-all,classicalSmatrix1,classicalSmatrix2,classicalSmatrix3}. Interestingly, we find that the NSDI features of the contact interaction from the 3-d calculations can even be captured with a 2-d semi-classical calculation, as shown in Figs. \ref{fig.tunneling}(c) and (d).

However, it is not clear whether these distinct features, which have been observed in the $S$-matrix results, are exclusively related to the range of the e-e interaction (long range versus short range or zero range). For example, they might also be artifacts of the assumptions made in the SFA. Also, it is of interest whether or not they are of quantum-mechnical origin, in the sense that they are dependent on the first electron being born by tunneling.  To clarify these issues, we perform NSDI calculations using the classical ensemble method \cite{Panfili-etal01}. Our results suggest that the different features are the combined effect of the SFA and the form of e-e correlation used in the $S$-matrix and semi-classical calculations and can be reproduced without tunneling.

Our paper is organized as follows.  In Sec. II, we first briefly introduce our two-electron (2-e) classical model and the computational method.  In Sec. III, this classical method is used to compute the final momentum distributions of NSDI electrons via either a long-range or a short-range e-e repulsion potential, and these classical distributions are compared with the corresponding $S$-matrix results.  In Sec. IV, we present a quantitative analysis of the recollision mechanism that leads to NSDI for both the long-range and the short-range e-e interaction.  In Sec. V, we identify the distinct classical NSDI trajectories and analyze their individual contribution to the final momentum distributions.  We then examine the effect of the SFA on our classical results and relate them to the S-matrix results in Sec. VI.  A summary is then presented in Sec. VII.

\section{Classical Model}

In our classical model, the Hamiltonian of a two-electron atom in an intense laser field, described in the long-wavelength approximation by its electric field $\vec{E}(t)$, is given by
\begin{equation}
H=\sum_{i=1}^{2} \left(
\frac{|\vec{p}_i|^{2}}{2}+V_{\mathrm{n},i}(|\vec{r_i}|)+\vec{r}_i
\cdot \vec{F}(t)
\right)+V_\mathrm{ee}(r_{12}),
\label{eq.2ehamiltonian}
\end{equation}
where $V_{\mathrm{n},i}$ are  the nuclear binding potentials experienced by the two electrons $(i=1,2)$, $V_\mathrm{ee}$ is the electron-electron (e-e) repulsion potential and $r_{12}=|\vec{r}_1-\vec{r}_2|$ is the distance between the two electrons.  This Hamiltonian allows tracking  the electron pair throughout the laser pulse under the simultaneous influence of the laser field and the nuclear potentials.  Previously, we have successfully used the aligned-electron version of this 2-e model to reproduce the features of the experimental NSDI ion-count data and the momentum distributions \cite{Ho-etal05}.   In this paper, we employ both 2-d and 3-d classical-ensemble methods to examine the consequences of the range of the e-e correlation (long range vs. short range) and the effects of the SFA on the NSDI dynamics.  Mostly, the 2-d results will be  presented.

In this method, one million or more Newtonian two-electron (2-e) trajectories that follow the Hamiltonian (\ref{eq.2ehamiltonian}) are used to obtain the classical response of a single 2-e atom to the influence of a linearly polarized intense laser pulse.   Our 780 nm ($\omega=0.0584$ a.u.) laser pulse has a peak intensity of 0.3 PW/cm$^{2}$ and a trapezoidal envelope with a symmetric 2-cycle turn-on and turn-off and a 4-cycle plateau. Prior to turning on the field, the electrons' positions and momenta form a micro-canonical ensemble with an energy of $-2.50$ a.u. and zero total angular momentum.

\section{Final Momentum Distributions}

To illustrate the features of and the differences between the long-range and short-range e-e interactions in NSDI, we perform several calculations using Coulombic and Yukawa e-e repulsion potentials, with each electron subjected to the same soft-core Coulombic nuclear potential \cite{Javanainen-etal87}, which has the familiar form
\begin{equation}
V_{\mathrm{n},i}(r_i)=\frac{-2}{\sqrt{r_i^2+a^2}}. \label{vni}
\end{equation}
We use $a=1.0$ to avoid auto-ionization in our ensemble of classical trajectories in the absence of the laser field.  In order to simulate the experimental data, we look for the characteristics of a long-range versus a short-range e-e interaction by examining the end-of-pulse NSDI longitudinal electron momentum distributions, which have all the transverse momenta integrated over, and relating them to the dynamical details of the associated double-electron ejection mechanism.

\begin{figure}
\centerline{\includegraphics[width=3.3in]{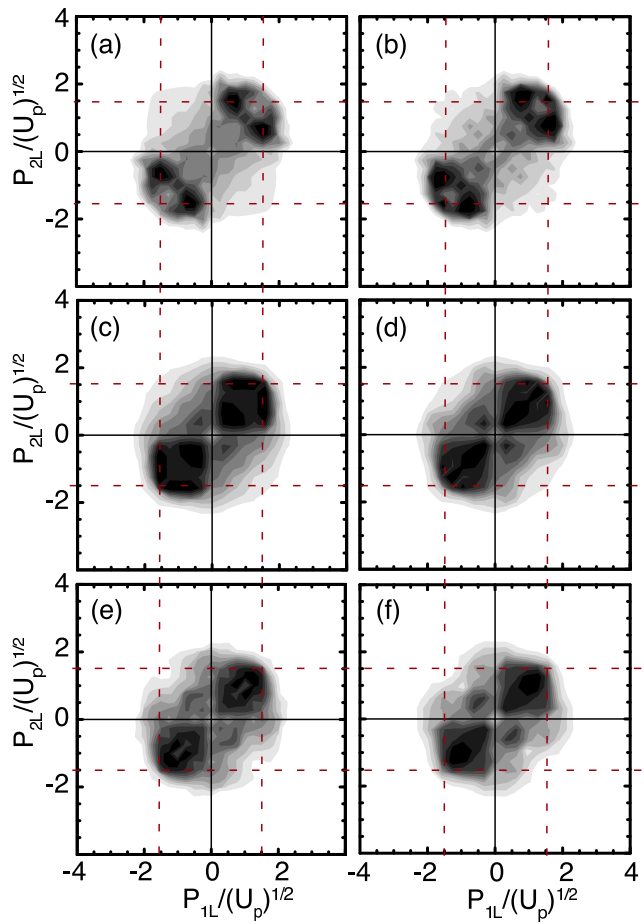}}
\caption{Two-electron longitudinal momentum distributions for NSDI obtained from the classical ensemble method at $I=0.3$ PW/cm$^2$. Panels (a) and (b) are calculated with a 3-d and 2-d soft-core Coulomb e-e repulsion potential respectively.  The remaining panels are calculated with 2-d Yukawa e-e repulsion potentials for different $\lambda$ values of (c) 0.5, (d) 2.0, (e) 4.0 and (f) 8.0. The vertical and horizontal (red) dashed lines mark P$_{1L}=\pm 1.5 \sqrt{U_p}$ and P$_{2L}=\pm 1.5\sqrt{U_p}$.  In each plot, the transverse momenta have been integrated over.  Each distribution is plotted on a linear gray scale with the darker regions representing higher density.}
\label{fig.2epdist}
\end{figure}

Figure~\ref{fig.2epdist}(a) shows the 2-e longitudinal momentum distribution obtained with a 3-d classical calculation using a soft-core Coulombic e-e repulsion potential with the form
\begin{equation}
V_\mathrm{LR,ee}(r)=\frac{1}{\sqrt{r^2+b^2}}
\label{SCrepulsion}
\end{equation}
with $b=0.1$.  First, we identify four features of this distribution and, below, we will examine how they are modified in the limit of a short-range e-e interaction:  (i) the NSDI momenta are largely confined to the region where $|p_{1L}|\alt 2\sqrt{U_p}$ and $|p_{2L}|\alt 2\sqrt{U_p}$, with  $U_p$ is the peak ponderomotive energy of an electron in the laser field;  (ii) the majority of the NSDI population is concentrated in the first and third quadrants, implying that with high probability the two NSDI electrons are emitted to the same side of the nucleus along the laser polarization axis;  (iii) there are nonnegligible populations in the second and fourth quadrants, which represent  NSDI events with opposite-side double-electron emission;  (iv) this distribution has well developed maxima on either side of the diagonal $p_{1L}=p_{2L}$, such that either $|p_{1L}|\approx 1.5 \sqrt{U_p}$ and $|p_{2L}|\approx 0.75 \sqrt{U_p}$ or $|p_{1L}|\approx 0.75 \sqrt{U_p}$ and $|p_{2L}|\approx 1.5 \sqrt{U_p}$.  Note that these four momentum features can be captured in detail both qualitatively and quantitatively with a reduced-dimensional 2-d calculation using the same form of the e-e potential in Eq.~(\ref{SCrepulsion}), as shown in Fig. \ref{fig.2epdist}(b).

These classical predictions from the long-range potential (\ref{SCrepulsion}) are consistent with the COLTRIMS measurements in He \cite{COLTRIMS-He}, Ne \cite{COLTRIMS-Ne}, and Ar \cite{COLTRIMS-Ar} in the intensity regime of NSDI.  In particular, like our classical distributions, all these experiments recorded electron pairs in all four quadrants, with the fractions in the first and third quadrants being dominant.  Also, dominant off-axis population [feature (iv) above] has so far been observed in experiments on He \cite{COLTRIMS-He} and Ar \cite{COLTRIMS-Ar-repul}. We will show that this off-axis population  is the signature of the Coulombic or long-range e-e repulsion. It disappears in the limit of a short-range e-e repulsion. Feature (iv)  has also been seen in  1d simulations, both in quantum \cite{lein} and in classical \cite{ho-etalPRA} calculations.

To illustrate this, we perform different classical calculations using a Yukawa potential, which has the form
\begin{equation}
V_\mathrm{SR,ee}(r)=\frac{\exp[-\lambda
\sqrt{r^2+b_1^2}]}{\sqrt{r^2+b^2}}.\label{Yukawa}
\end{equation}
Note that in the limit of $\lambda=0$, this Yukawa potential is reduced to Eq.~(\ref{SCrepulsion}).  Since there is excellent agreement between the 3-d and 2-d calculations using a soft-core potential, we restrict our calculations with the Yukawa potential to 2-d. With $b_1=b=0.1$, Figs.~ \ref{fig.2epdist}(c) and (d) show that, at $\lambda=0.5$ and 2.0, the momentum distributions still possess remnants of the long-range repulsion signature (iv). As a new feature, there is a high concentration of electron pairs with momenta around $|p_{1L}|\approx |p_{2L}|\approx 0.7 U_P$.  The presence of these low-momentum electron pairs can be regarded as a signature of an intermediate-range e-e interaction.

In order to reach the limit of a short-range interaction, the value of $\lambda$ in the Yukawa potential (\ref{Yukawa}) must be further increased. However, calculations with  large values of $\lambda$ require significantly increased  computation time.  Fortunately, we find that the momentum distribution calculated with $\lambda=4$ already simulates the features of a short-range e-e interaction sufficiently well. This is due to the fact that, at $\lambda=4$, the interaction energy of the electron pair drops by more than a factor of 50 when the two electrons move merely 1 a.u. away from each other. Furthermore, comparison between Figs. \ref{fig.2epdist}(e) and (f) shows that raising the value of $\lambda$ from 4 to 8 does not modify the momentum distribution significantly.  Thus, it is sufficient to extract the classical features of a short-range e-e interaction from calculations using the Yukawa potential with $\lambda=4$.

Comparison between Figs. \ref{fig.2epdist}(b) and (e) shows that one can identify two tendencies as the e-e interaction is changed from long range to short range.  First, as expected, the Coulombic repulsion signature (iv) disappears and the regions with the highest density of NSDI events are closer to the diagonal $p_{1L}=p_{2L}$ and near $|p_{1L}|=|p_{2L}|=1.5 \sqrt{U_p}$.  Second, in the regions near the origin, especially in the 2nd and 4th quadrants, the population density  becomes higher.  Otherwise, the location of the cutoffs and the relative population distribution in the four quadrants remain largely unchanged.

Before we examine the dynamics of the NSDI events in both long-range and short-range interactions, it is interesting to observe that our classical calculations capture some essential features of the $S$-matrix and the semi-classical calculations, even though they do not include tunneling and go beyond the SFA.  Comparison of Figs.~\ref{fig.tunneling} (a) and (b) on the one hand with Figs.~\ref{fig.2epdist} (b) and (f) on the other  shows rather good agreement. Especially, in the case of the long-range e-e interaction, the most probable NSDI events have one fast and one slow electron, and we will show later that our classical analysis suggests that the recolliding electron is likely to be the slow electron.  Apart from these agreements, there are two noticeable differences between our classical predictions and the $S$-matrix results.  First, our classical momentum distributions exhibit a higher population in the second and fourth quadrants regardless of the form of the e-e interaction. Second, as the e-e interaction changes from Coulomb to contact interaction, the $S$-matrix distributions become much more extended in the first and third quadrants, and there is less population near the origin in all four quadrants.  The role of the SFA and of tunneling in this context will be examined below.

\section{Travel times and thermalization times}

To understand in detail the NSDI dynamics caused by long-range or short-range e-e potentials, we follow individual NSDI events in both potentials throughout the laser pulse.  We find that the rescattering mechanism without tunneling continues to drive NSDI  for both potentials.  For each NSDI event, we determine three times that characterize its temporal evolution: the time $t_\mathrm{si}$ of the single-ionization event, the time $t_\mathrm{recol}$ of the recollision, and the time $t_\mathrm{NSDI}$ of double ionization. Note that such characterization is in line  with the three-step recollision model underlying the $S$-matrix and semi-classical calculations.  For our classical simulations, a completely accurate determination of these times is not possible since they have to be extracted during the course of non-equilibrium and chaotic 2-e dynamics.  Here, we adopt the following working definitions.  We define $t_\mathrm{si}$ as the time when one of the electrons first has its kinetic energy greater than the combined energy of its nuclear and e-e interactions.  Hence, $t_\mathrm{si}$ marks the ``time of birth" of the electron that later will recollide with the ion. The second time $t_\mathrm{recol}$ we define as the time of closest approach between the recolliding and the bound electron.  The double-ionization time $t_\mathrm{NSDI}$ is defined as the time when both electrons first have their individual kinetic energies greater than their nuclear potential energies.

\begin{figure}
\centerline{\includegraphics[width=3.0in]{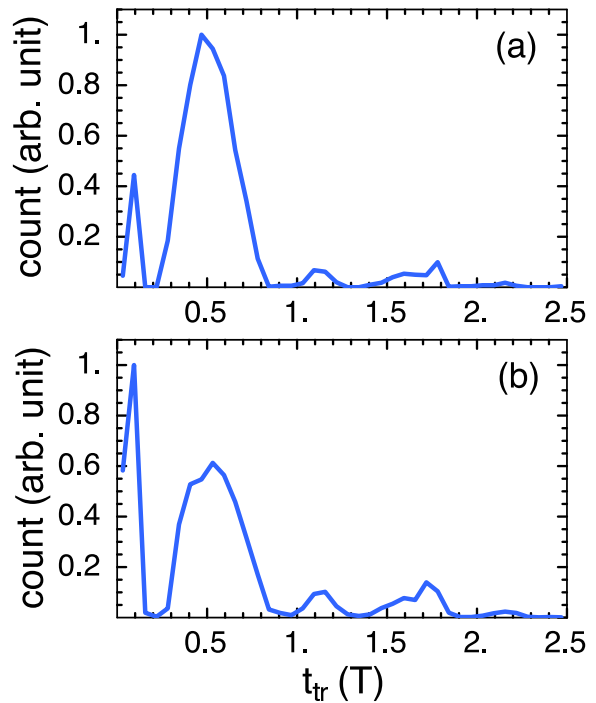}}
\caption{Travel time distribution for the recolliding electron of all NSDI trajectories obtained from the 2-d classical ensemble method with (a) long-range and (b) short-range e-e repulsion potentials. Each distribution is scaled to unity at the highest count.  }
\label{fig.traveltime}
\end{figure}

With these three different times, we can statistically quantify the NSDI processes in our 2-d classical ensemble of atoms in both e-e potentials.  Similar quantification schemes with different working definitions have been used to obtain the individual distributions of $t_\mathrm{recol}$ and $t_\mathrm{NSDI}$ in connection with the laser phase \cite{ Ho-etalNSDIpulselength, Haan-etalNSDIdelay}.  Here, we examine the distribution of two consecutive time intervals, $t_\mathrm{recol} - t_\mathrm{si}$ and $t_\mathrm{NSDI} - t_\mathrm{recol}$, among the NSDI trajectories, in order to obtain the time scales associated with the recollision mechanism.

The difference between $t_\mathrm{recol}$ and $t_\mathrm{si}$  is the time interval that the rescattering electron travels outside the vicinity of the nucleus before returning to the nucleus for the recollision, in short the ``travel time''.  Figures \ref{fig.traveltime}(a) and (b) show that the probability distributions of the travel time $t_\mathrm{recol}-t_\mathrm{si}$ of all NSDI trajectories obtained from the 2-d soft-core Coulombic e-e interaction and the Yukawa e-e repulsion potential with $\lambda=4$ are very similar: both distributions are concentrated within certain well-defined ranges of the travel time. These are centered about  the travel times $t_\mathrm{tr}=0.1,\,0.5,\,1.15,\,1.75,$ and $2.2\,T$, where $T=2\pi/\omega$. The classical simple-man model (or, equivalently, the SFA) yields the values $t_\mathrm{tr}=0.65,\,1.2,\,1.7$, and $2.2\,T$; see, e.g. Ref.~\cite{PKGB02}. The close agreement between the travel times calculated from the present classical 2-d simulation, which contains all interactions,  and from the simple-man model, which ignores the nuclear potentials, is very remarkable.  It is interesting to observe that the complex saddle-point approximation to the SFA, like the full classical simulation shown here, also yields a solution with a very short travel time  $(t_\mathrm{tr}\approx 0.1T)$, which exclusively contributes to harmonic generation with harmonic order below the first ionization energy \cite{MB02}.   The close agreement between the present calculations with zero-range, short-range, and long-range potentials implies that the travel times of the rescattering electron are potential independent and their well-defined ranges are mainly determined by the interaction with the laser field.    Among the classical NSDI events, those with travel times shorter or longer than $T$ can be regarded as single- or multiple-recollision events, respectively.   In particular,  of the NSDI events mediated by the long-range and the short-range interaction,  more than $90\%$ and $80\%$, respectively, are single-recollision events.  A detailed analysis of the single-recollision events will be presented later in the paper.

\begin{figure}
\centerline{\includegraphics[width=3.0in]{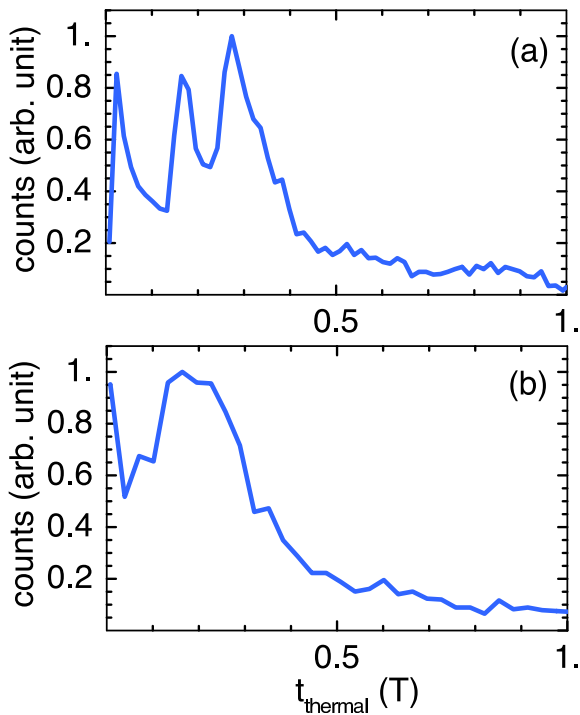}}
\caption{Thermalization time distribution for the recolliding electron of all NSDI trajectories obtained from the 2-d classical ensemble method with (a) long-range and (b) short-range e-e repulsion potentials. Each distribution is scaled to unity at the  highest count.}
\label{fig.therm-time}
\end{figure}

The second time interval, $t_\mathrm{NSDI} - t_\mathrm{recol}$, is the time delay between double ionization and recollision.  This quantity is sometimes interpreted as the ``thermalization time'', $t_\mathrm{thermal}$, i.e. the time that is needed for both electrons to acquire enough energy from the recollision process such that they can liberate themselves from the nucleus \cite{Liu-thermalization}. Even though the terminology refers to a large number of electrons we use it here for an ensemble of just two electrons.  Figures \ref{fig.therm-time} (a) and (b) show that both long-range and short-range e-e interactions generate some NSDI events with very short or close to zero thermalization time. Previously studied using quantum $S$-matrix and simple-man approaches \cite{s-matrix-bf, s-matrix-mbi, s-matrix-gp, s-matrix-all, classicalSmatrix1, classicalSmatrix2, classicalSmatrix3} tacitly assumed zero thermalization time.  However, Fig.~\ref{fig.therm-time} shows that most of the NSDI trajectories exhibit a longer thermalization time, even as long as $4.0\,T$ (not shown in Fig. \ref{fig.therm-time}), for both forms of the e-e potential.   In the case of the long-range e-e interaction, these events develop two local maxima near $0.15\,T$ and $0.25\,T$.  Similarly, the distribution of $t_\mathrm{thermal}$ for the short-range e-e potential has one broad dominant peak near $0.2\,T$. Liu et al. performed a semiclassical statistical simulation \cite{Liu-thermalization} of  the experimental ion-momentum distributions of triple and quadruple nonsequential ionization of neon \cite{multi,multi2}. They obtained a best fit for $t_\mathrm{thermal} \approx 0.17\,T$, in close agreement with the range between  $0.15\,T$ and $0.25\,T$ displayed in Fig.~\ref{fig.therm-time}.  Regardless of the form of the e-e potentials, only less than $25\%$ of the NSDI events have $t_\mathrm{thermal} > 0.5\,T$.  These long-$t_\mathrm{thermal}$ events contribute significantly to the production of electron pairs with opposite longitudinal momenta and ions with low longitudinal momentum \cite{Ho-etal05,Haan-etalNSDIdelay,Ho-etal03}.

\section{Long and short travel time trajectories}

As mentioned earlier, both long-range and short-range e-e interactions are very effective in producing NSDI with only one recollision.  For the further analysis, we subdivide these single-recollision events into short-travel-time (STT)  and long-travel-time (LTT) trajectories, according to their travel times being shorter or longer than $0.2 T$.  Figure ~\ref{fig.traveltime} shows that both the long-range and the short-range e-e interaction produce more LTT than STT trajectories.  But the short-range interaction leads to more STT trajectories (more than 25\%) than the long-range interaction (fewer than 20\%).

\begin{figure}
\centerline{\includegraphics[width=3.3in]{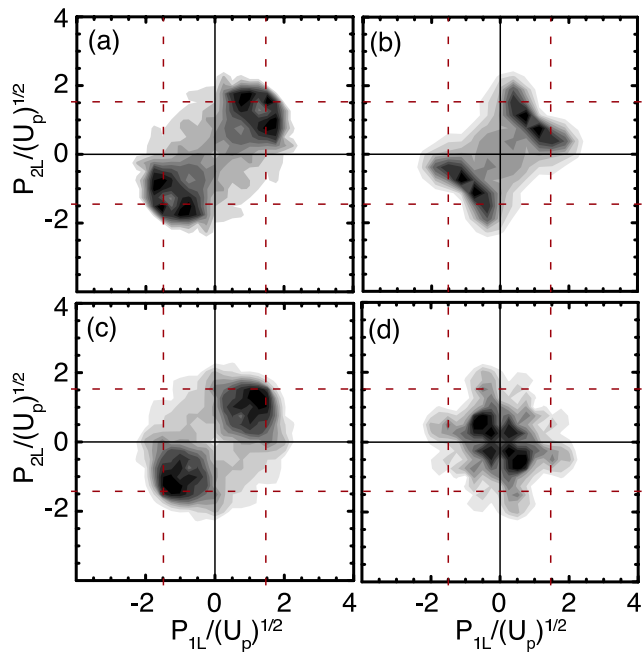}}
\caption{Classical 2-e longitudinal momentum distributions of two sub-ensembles of NSDI trajectories: LTT (left column) and STT (right column) trajectories. The panels in the top and bottom rows are obtained from the soft-core Coulomb e-e repulsion potential and the Yukawa e-e repulsion potential, respectively. The vertical and horizontal (red) dashed lines mark P$_{1L}=\pm 1.5 \sqrt{U_p}$ and P$_{2L}=\pm 1.5\sqrt{U_p}$.  In both plots, the transverse momenta have been integrated over.  The distributions are plotted on a linear gray scale with the darker regions representing higher density.}
\label{fig.LTTSTT2epdist}
\end{figure}

\begin{figure}
\centerline{\includegraphics[width=3.3in]{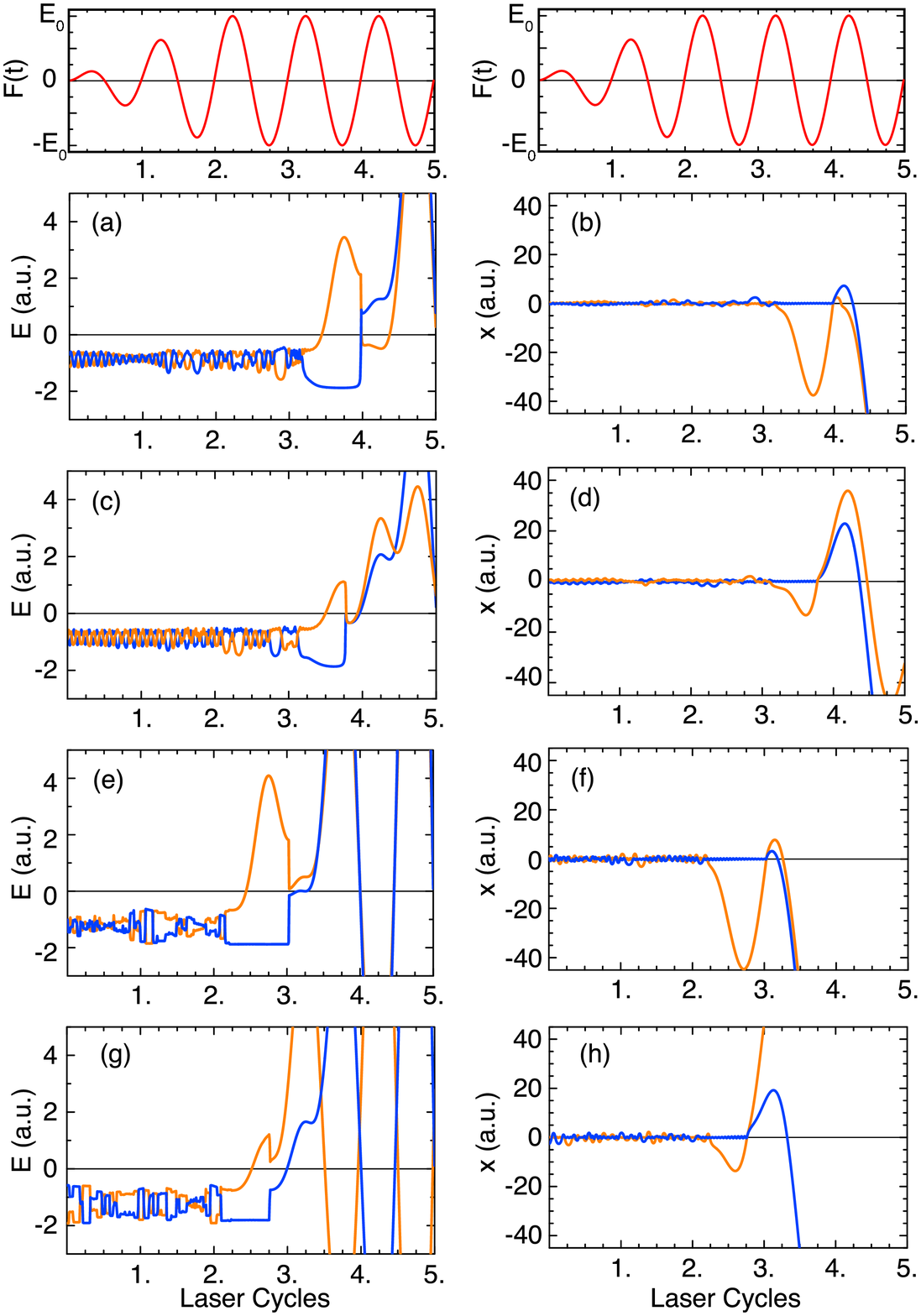}}
\caption{(Color online) Time evolution of the energy and the longitudinal displacement of NSDI electrons during the first 5 laser cycles. The top panel of each column plots the time evolution of the laser electric field   $F(t)$ with the peak amplitude $E_0$. The remaining panels of each column display the energy $E$ and the transverse displacement $x$ of the two NSDI electrons of a trajectory, respectively. The second and third row correspond to a representative LTT and STT trajectory, respectively, obtained with the soft-core Coulombic e-e repulsion potential, the fourth and fifth to the same for the Yukawa e-e repulsion potential.  Note that the energy of each electron is the sum of its kinetic, nuclear binding, e-e repulsion, and laser-dipole energies. The curves referring to the recolliding and the bound electron are painted orange (grey) and blue (black), respectively. }
\label{fig.traj}
\end{figure}

Analysis of our trajectories shows that the final momenta of the electron pairs of LTT and STT trajectories are quite different. Figures~\ref{fig.LTTSTT2epdist}(a) and (b) show that, regardless of the form of the potential, significantly nonzero momenta are almost exclusively generated by LTT trajectories. For either interaction, the contribution of the STT trajectories is relatively minor. For the long-range interaction, it is insignificant for all momenta; cf. Fig.~\ref{fig.2epdist} (a) on the one hand and Figs.~\ref{fig.LTTSTT2epdist} (a) and (b) on the other. For the short-range interaction, the STT trajectories do make important contributions for low momenta.

Dynamically, LTT and STT trajectories are quite different. Figure~\ref{fig.traj} demonstrates their differences with trajectories that have both electrons ejected and recolliding within the same laser cycle.  Here, we point out three differences among the LTT and STT trajectories that hold for both the long-range and short-range e-e interaction.  First, during recollision the energy carried by the rescattering electron of an LTT trajectory is larger than that of an STT trajectory.  In particular, the energy of the rescattering electron of an LTT trajectory can be higher than the ionization potential of its bound electron [see Figs.~\ref{fig.traj}(a) and (e)], while this is not the case for an STT trajectory [see Figs.~\ref{fig.traj}(c) and (g)].   Second, the recollision of an LTT trajectory can take place over a wide range of laser phases, especially near zero field strength, and still free both electrons. On the other hand,  for an STT trajectory, the time of the recollision is mainly restricted to laser phases where the field strength is near its maximum.  This implies that laser suppression of the nuclear potential barrier is important in leading to the double-electron ejection. Third, after recollision, both the bound and the rescattering electrons of an LTT trajectory tend to execute a short longitudinal excursion in the direction of incidence of the rescattering electron and are likely to move away together in the opposite direction of incidence. The STT trajectories, on the other hand, are likely to have both electrons executing a rather long longitudinal excursion in the direction of incidence of the rescattering electron.  Subsequently, both electrons tend to drift away together in the direction opposite to the direction of incidence if the e-e interaction is long range, but in opposite directions if it is short range.

\section{Effects of the SFA}

So far, we have focused on the different features of NSDI mediated by a long-range or a short-range e-e interaction while both electrons are continuously subject to the Coulombic nuclear potential.  However, it is not clear how robust these features are to the form and influence of the nuclear potential.  This is especially important because the SFA ignores the nuclear potential after an electron has become free \cite{footnote2}. In particular, we want to know the effect of suppressing the recolliding electron's interaction with the nucleus in NSDI processes, one though not the only approximation made by the SFA.  Here, we will not artificially enforce the second approximation of the SFA, where the laser interaction of the bound electron is ignored.  We can do that because the bound electron of each classical trajectory is bound so deeply in its nuclear potential well that the laser field practically has no effect on it.  This is evident from Fig.~\ref{fig.traj}, which shows that the energy of the bound electron of different trajectories is rather constant during the time interval between the birth of the rescattering electron and its recollision.

Thus, in order to mimic the SFA, we construct an explicitly time-dependent nuclear potential for each electron by replacing the electron-nucleus interaction potential (\ref{vni}) according to $(i=1,2)$
\begin{equation}
V_{\mathrm{n},i}(r_i) \rightarrow
V_{\mathrm{n},i}(r_i,t)=f_i(t)\frac{-2}{\sqrt{r_i^2+a^2}},
\end{equation}
with $a=1$.  If the rescattering electron, say electron $\#$1, is first ionized and has satisfied the supplementary condition of having moved more than 10 a.u. away from the core at $t=t_\mathrm{0}$, then we turn off its interaction with the nucleus according to
\begin{eqnarray}
\label{SFA}
f_1(t) &=&\left\{ \begin{array}{ll} 1, & \mbox{if $t \le t_\mathrm{0}$} \\
     e^{-\mu (t-t_\mathrm{0})/T}, & \mbox{if $t > t_\mathrm{0}$}
     \end{array}
                                                              \right. \\
f_2(t)&=&1,  \nonumber
\end{eqnarray}
where $\mu=40$ and $T$ is the laser period. In the limit when $\mu=0$, this nuclear potential is reduced to the soft-core potential.  These factors of $f_1(t)$ and $f_2(t)$ and our chosen value of $\mu$ ensure that once the rescattering electron leaves the nucleus and begins its longitudinal excursion, its binding potential is turned off smoothly and exponentially in time, whereas the form of the nuclear potential for the bound electron remains unaltered even after double ionization \cite{footnote1}.

\begin{figure}
\centerline{
\includegraphics[width=3.3in]{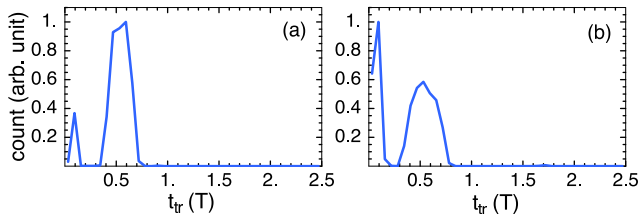}}
\caption{Travel time distribution for the recolliding electron of all NSDI trajectories obtained from the 2-d classical ensemble method under the SFA with (a) long-range and (b) short-range e-e repulsion potentials.}
\label{fig.SFAtraveltime}
\end{figure}

Note that our definition of $t_\mathrm{0}$ is to ensure that we begin turning off the nuclear potential of the rescattering electron only when its nuclear attraction has become small, i.e. when it is far away from the nucleus.  Here we have chosen the distance of 10 a.u. for the supplementary condition, because this distance is less than half of the jitter radius of the electron, which is $\sqrt{I}/\omega^2 \approx$ 27 a.u., in a laser field with $I=0.3$ PW/cm$^2$.  Thus, this supplementary condition provides us with a sufficiently long time window before the time of recollision to turn off the nuclear interaction smoothly in time. Furthermore, with $\mu=40$,  we can ensure that the nuclear interaction of the rescattering electron is turned off promptly and effectively long before the recollision takes place.  Failure to turn off the nuclear interaction of the rescattering electron smoothly in time prior to recollision can produce a large number of unrealistic NSDI events that involve self-ionization \cite{self-ionized}.

An  immediate consequence of the SFA as expressed in the turned-off potential (\ref{SFA}) is that it eliminates NSDI events caused by LTT trajectories with a travel time longer than one cycle, cf. Figs. \ref{fig.traveltime} and \ref{fig.SFAtraveltime}. This is so because the nuclear potential (\ref{vni}) is essential to refocusing the first-ionized electron towards the nucleus. In its absence, due to the electron-electron repulsion, it will be deflected for good and not return anymore to the vicinity of the ion \cite{ho_eberlyNSTI}. Comparison of Figs.~ \ref{fig.traveltime} and \ref{fig.SFAtraveltime} also shows that the travel time of the first recollision event  can be regarded as a rather robust feature both for the long-range and for the short-range e-e interaction, since the individual patterns are only altered slightly when the rescattering electron's nuclear interaction is turned off exponentially in time.

\begin{figure}
\centerline{
\includegraphics[width=3.3in]{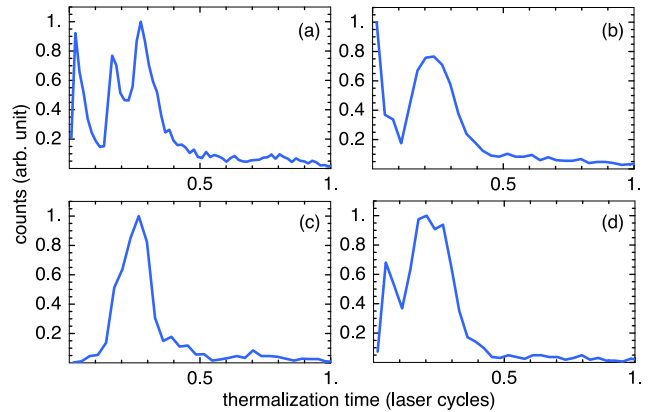}}
\caption{Thermalization time distribution for the recolliding electron
of the LTT NSDI trajectories obtained with the soft-core Coulombic
(left column) and the Yukawa (right column) e-e repulsion potential
without (top row) and with (bottom row) the SFA.   Each distribution
is scaled to unity at the highest count.}
\label{fig.LTTtherm-time}
\end{figure}

Despite the fast turn-off of the nuclear interaction according to Eq.~(\ref{SFA}), the rescattering electron of the STT trajectories may still experience a non-negligible nuclear interaction prior to the recollision for a large fraction of its travel time.  Hence, in order to clearly demonstrate the effect of the SFA, we need to choose the sub-ensemble of the LTT trajectories. They have a travel time longer than $0.2T$, so that their rescattering electrons spend more than half of their travel time with the strength of their nuclear interaction suppressed by more than a factor of 50.  In particular, we demonstrate the effect of the SFA on the thermalization-time distribution and the end-of-pulse electron momentum distribution for the LTT trajectories.

Figure \ref{fig.LTTtherm-time} shows that under the SFA the production of LTT trajectories with a very short thermalization time is suppressed significantly regardless of the form of the e-e potential. Apart from this, the distributions obtained with and without the SFA display similar characteristics.  First, there are relatively few trajectories with thermalization times longer than $0.5\,T$. Second, the location of the local maximum in each distribution is in the range of $0.15$ to $0.25\,T$.  These similarities combined with the similar range of travel times (cf. Fig.~\ref{fig.SFAtraveltime}) indicate that for the LTT trajectories the SFA only slightly modifies the recollision scenario and its time scales.

\begin{figure}
\centerline{\includegraphics[width=3.3in]{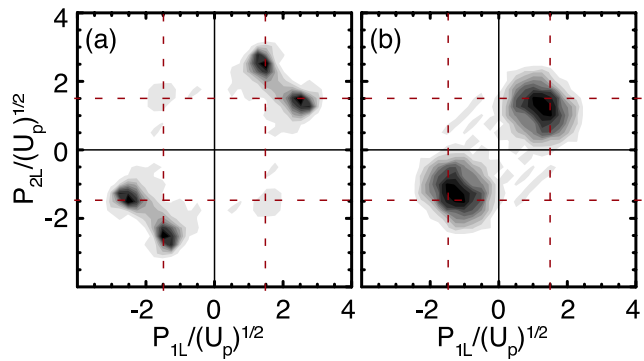}}
\caption{Classical 2-e longitudinal momentum distributions under the SFA for LTT trajectories obtained from (a) the soft-core Coulomb e-e repulsion potential and (b) the Yukawa e-e repulsion potential. The vertical and horizontal (red) dashed lines mark $P_{1L}=\pm 1.5 \sqrt{U_p}$ and $P_{2L}=\pm 1.5\sqrt{U_p}$.  In both plots, the transverse momenta have been integrated over.  Panels (a) and (b) are to be compared with Figs. \ref{fig.LTTSTT2epdist} (a) and (c), respectively.}
\label{fig.SFA2epdist}
\end{figure}

Having only slight modifications in the recollision process, however, does not guarantee that the classical calculations with the SFA should reproduce the essential features of the final-state momentum distributions from the corresponding calculations without the SFA.  Figures~\ref{fig.LTTSTT2epdist}(a), \ref{fig.LTTSTT2epdist}(c), \ref{fig.SFA2epdist}(a), and \ref{fig.SFA2epdist}(b) allow us to compare the momentum distributions of the LTT trajectories that are obtained, roughly speaking, with and without the SFA.  The comparison shows that the SFA has a dramatic effect on NSDI events caused by the long-range e-e interaction; cf. Figs.~\ref{fig.LTTSTT2epdist}(a) and \ref{fig.SFA2epdist}(a). The SFA produces NSDI electrons with higher momenta such that the boundary of the distribution is pushed beyond $ |p_{1L}| = |p_{2L}| = 2.5\sqrt{U_p} $ while the population in the region where both $|p_{1L}|$ and $|p_{2L}|$ are smaller than $1.5\sqrt{U_p}$ is strongly diminished. Also, while the majority of the population remains in the 1st and 3rd quadrants, the SFA leads to one additional change, namely the highest-population-density regions become well separated, and each region is localized and moves away from the diagonal $p_{1L}=p_{2L}$. This implies that NSDI events under the SFA are more likely to produce one fast and one slow electron.  This is reminiscent of the characteristics of the distribution obtained via $S$-matrix and semi-classical calculations with the Coulombic e-e interaction, even though our results are not identical with those.

The SFA also modifies the appearance of the momentum distribution from the short-range e-e interaction; cf. Fig.~\ref{fig.SFA2epdist}(b). Even though the modification is noticeable, it still retains the main features of the short-range interaction without the SFA.   For instance, the cutoffs are still near $ |p_{1L}| = |p_{2L}| = 2.0\sqrt{U_p} $ and the regions with the highest population density are still centered near $ p_{1L} = p_{2L} = \pm 1.5 \sqrt{U_p} $.  The noticeable change is that there is now relatively little population near the origin, on both the $x$ and the $y$ axis and in the 2nd and 4th quadrant. Surprisingly, this classical distribution under the SFA adequately mimics the overall features of the distribution obtained via $S$-matrix and semi-classical calculations using a contact interaction.  This good match suggests that these features have a classical origin and are not dependent on the electron being born by tunneling.

\begin{figure}
\centerline{\includegraphics[width=3.3 in]{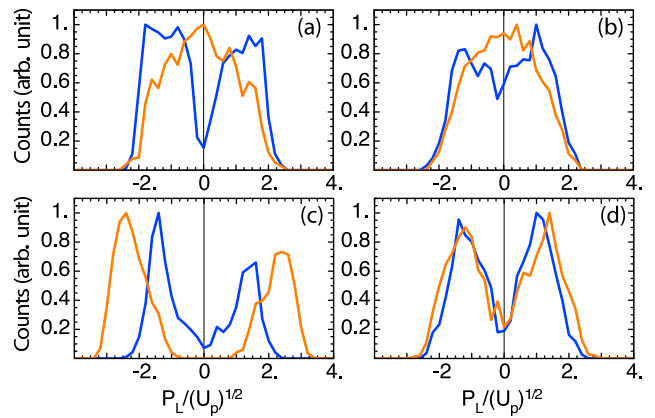}}
\caption{Momentum distributions of the rescattering and bound electrons in LTT trajectories obtained with the soft-core Coulombic (left column) and the Yukawa (right column) e-e repulsion potential.  The panels in the top and bottom rows are obtained without and with the SFA, respectively. The orange (gray) and blue (black) lines correspond to the momentum distribution of the rescattering and the bound electron, respectively.}
\label{fig.SEpdist}
\end{figure}

With our classical ensemble method, one can further determine the effect of the SFA directly on the rescattering and the bound electron by examining their final momentum distributions for all LTT trajectories. Due to the strong coupling between the two electrons, turning off the recolliding electron's nuclear potential has a substantial effect not only on this electron but also on the bound electron, as is shown in Fig. \ref{fig.SEpdist}.   For the long-range e-e interaction, the distributions of  the rescattering and the bound electron have a similar range of final longitudinal momenta, extending from about $-2\sqrt{U_p}$ to $2\sqrt{U_p}$, when the nuclear interaction of the rescattering electron is present at all times.  But the two distributions have different shapes. The distribution of the rescattering electron peaks at zero momentum, whereas the distribution of the bound electron electron has a double-peak structure with a deep valley at zero momentum. Thus, we can roughly label the rescattering and the bound electron as the slow and the fast electron, respectively.  However, Fig.~ \ref{fig.SEpdist}(c) shows that the SFA permutes these labelings. Both electrons now have a double-peak structure. the rescattering electron has become the faster electron,  and its momentum range  has expanded to between $-3\sqrt{U_p}$ and $3\sqrt{U_p}$, with very little population of momenta smaller than $\sqrt{U_p}$ \cite{anotherfootnote}. Momenta this high can only be the  result of the rescattering electron being backscattered by the bound electron and thereafter acquiring momentum from the laser field \cite{anotherfootnote}.   As for the bound electron, its momentum distribution under the SFA also has a pronounced double-peak structure with a relatively small fraction of its population having near zero momentum. However, its range is still limited to between $-2\sqrt{U_p}$ and $2\sqrt{U_p}$.


The situation, however, is rather different for the short-range e-e interaction. With and without out the SFA, the momenta of both electrons are between  $-2\sqrt{U_p}$ and $2\sqrt{U_p}$, but the distributions are not identical; the rescattering electron is more likely to have a small momentum than the bound electron.  Under the SFA, the range of the momenta of both electron remains the same,  but the momentum distributions of both electrons become almost identical, exhibiting pronounced valleys around zero momentum. This  signals very efficient momentum sharing among the two during the recollision.  The absence under the SFA of high momenta  implies that the rescattering electron experiences little backscattering from the bound electron for the short-range e-e interaction.

It is remarkable that the effect of the SFA is much larger for the Coulombic than for the short-range e-e interaction. Without the SFA, the two potentials yield momentum distributions that are rather similar; cf. Figs. \ref{fig.SEpdist} (a) and (b). With the SFA, the Coulombic interaction produces very many electrons with unusually high momenta [Figs. \ref{fig.SEpdist} (c) and \ref{fig.SFA2epdist} (a)], which betray the typical Coulomb signature of one fast and one slow electron. In contrast, for the short-range e-e interaction, the effect of the SFA is much weaker and tends to produce electrons with  closely related momenta. For the short-range interaction, the momentum distribution of Fig.~\ref{fig.SEpdist} (d), which was obtained with the SFA, agrees quite well with the corresponding $S$-matrix and the semiclassical results.

\section{Summary}

In summary, we have carried out classical simulations of nonsequential double ionization with the electron-electron repulsion potential represented either by a soft-core Coulombic or by  a Yukawa potential. We have also mimicked the strong-field approximation, which is commonly made in $S$-matrix calculations, by gradually eliminating the electron-nucleus interaction when the first-ionized electron reaches a certain distance from the nucleus.  We identify a distinct signature of the long-range e-e interaction in the momentum distribution. Several features of our classical calculations can be observed in the  experimental data, depending on the atomic species on the one hand and the form of the e-e interaction on the other.

Unlike their effect in the quantum-mechanical $S$-matrix calculations, the soft-core Coulombic and the Yukawa potential produce fairly similar classical NSDI momentum distributions, as long as the  strong-field approximation is not invoked. The notable exception is that the short-range potential tends to produce more electrons with the same or almost the same longitudinal momenta. In addition, for both potentials the distributions of the travel time -- that is, the time between ionization and recollision -- are rather similar, and they are in close agreement with the prediction from the simple-man model, except for the existence of trajectories with very short travel times below $0.2T$, which are not provided by the simple-man model.  The latter  are more important in the case of the short-range than the long-range potential.  Furthermore, the most probable thermalization time  -- that is, the time between recollision and the electrons becoming free -- extracted from our classical calculations is within the range of 0.15 $T$ to 0.25 $T$. This is of the same order as the value that yielded the best agreement of a statistical model with experimental data for nonsequential multiple ionization.

Surprisingly, we find that our adaptation of the SFA can modify the momentum distributions substantially, while having little or no effect on the travel time distribution.  One common feature is that the SFA suppresses trajectories with long travel time when the longitudinal momenta of both electrons both are low or have opposite sign.  In the case of the short-range e-e interaction, we compare our classical simulations augmented by our SFA with the $S$-matrix and the semi-classical calculations that employ a contact interaction. The agreement of these results is quite good.  This implies that tunneling is not neccessary to generate the features from of the contact interaction.  For a long-range potential, the discrepancies between our classical and the $S$-matrix calculations are more pronounced.  These discrepancies can be due to the first-order Born approximation used in the $S$-matrix calculations.

\section*{Acknowledgments}
This work was supported by DOE Grant DE-FG02-05ER15713.  PJH acknowledges the summer fellowship from Deutscher Akademischer Austausch Dienst (DAAD) and discussions with J. H. Eberly.


\begin{thebibliography}{99}
\bibitem{footnote2} The short-range electron-electron interaction has been, in the zero-range limit, interpreted as a three-body interaction whose action is restricted to the position of the nucleus. Hence, in this sense, the short-range e-e interaction may reflect the presence of the nuclear potential. However, in the SFA it is only treated in the first-order Born approximation \cite{classicalSmatrix2}.
\bibitem{footnote1} One can also turn off the nuclear potential of the bound electron after double ionization, such that both electrons are free of the influence of the nucleus in their final state.  However, doing so only increases the final momentum of the bound electron slightly and does not affect our conclusion regarding the effect of the SFA.
\bibitem{self-ionized} In the case when the nuclear interaction of the rescattering electron is not turned off substantially at the time of recollision, it is possible that the rescattering electron can be recaptured by its weak but nonnegligible nuclear potential if it transfers a substantial fraction of its energy to free the bound electron.  Note that the nuclear interaction of the rescattering electron is not restored and is diminished continuously in time, even if it is recaptured.  Thus, shortly after the escape of the bound electron, the nuclear attraction of the recaptured rescattering electron will become sufficiently weak that it can no longer hold the recaptured electron and the electron will become ionized.
\end{thebibliography}
\end{document}